# EFFICIENT IMPLEMENTATION OF GALS SYSTEMS OVER COMMERCIAL SYNCHRONOUS FPGAS: A NEW APPROACH

Javier D. Garcia-Lasheras

garcialasheras@na2.es

Electrical and Electronic Engineering Dept., Public University of Navarre

## ABSTRACT

**This paper introduces a new approach for implementing Globally Asynchronous – Locally Synchronous (GALS) systems over commercial available Field Programmable Gate Arrays (FPGA). This new vision is aimed to overcome the logic overhead issues that previous works exhibit when applying GALS techniques to programmable logic devices. The proposed new view relies in a 2-phase / bundled data parity based protocol for data transfer and clock generation tasks. The ability of the introduced methodology for smart real-time delay selection allows the implementation of a variety of new methodologies for electromagnetic interference mitigation and device environment changes adaptation.**

**KEYWORDS:** GALS, FPGA

## 1. INTRODUCTION

From the appearance of integrated transistor in the electronic industry, digital systems performance has spectacularly advanced year after year. Progressive miniaturization of CMOS processes maintains this growth because smaller transistor sizes allow higher operating frequencies and the integration of more complex logic functions in the same silicon die. Unfortunately, nowadays this paradigm is changing. Transistors are now as tiny that their characteristics in terms of power consumption and delay are comparable to those associated to the connections between them. This situation forces designers to take special care when implementing metal interconnections. The problem is particularly critical when we consider the huge skew-free clock distribution networks that are implemented in complex chip designs. These networks not only consume an enormous amount of power, but also cause a harmful electromagnetic interference (EMI) that can affect their own correct operation and those of others systems surrounding them. In order to cope with this situation, engineers are focusing their attention towards the implementation of Globally

Asynchronous Locally Synchronous systems. In these specially designed systems, clock network is split into different synchronous clock domains with asynchronous circuitry controlling their correct coordination.

In the last years, as Field Programmable Gate Arrays are being implemented in deep submicron processes, we can extrapolate that issue to this special class of electronic devices. Contrarily to full-custom design, engineers trying to deploy GALS systems into commercially available FPGAs must deal with a fixed architecture that determines the different kinds of asynchronous circuitry that can be implemented in an efficient way. This situation drives irremediably to a logic resources overhead that precludes the use of GALS approach over these devices for commercial applications [1].

In this paper, we will try to overcome these issues by introducing a new GALS approach for programmable devices that is strongly inspired by the CMOS GasP [2] minimal control protocol for FIFO micropipelines [3]. In this way, we have designed a General Purpose Rendezvous Module (GPRM) aimed to match with the fixed architecture of commercial LUT-based FPGAs [4]. This module has added functionalities over the standard straight-forward FIFO control like clock pulse generation for self-timed operation of independent synchronous blocks and smart real-time cycle length selection for implementing EMI spread spectrum and throughput control techniques.

## 2. PROTOCOL IMPLEMENTATION

### 2.1. ARCHITECTURAL DESCRIPTION

As it is the case of any GALS approach, our methodology is based on the decomposition of the whole system into Autonomous Processing Blocks (APB). In Figure 1 we can see the general scheme of our proposed architecture by focusing into an individual APB. The main part of the block is composed by a registered logic function (LF). The value contained in the own APB register and those originated by external ones act as inputs for each logic function. We must note that the registered logic function will always be a conventional synchronous design. The most common way for implementing the LF is a placement of a HDL description synthesized by an appropriated EDA tool, but in the new generation of high density platform FPGAs the logic function can include some hard-IP like DSP slices, memory blocks or even full processors [4].

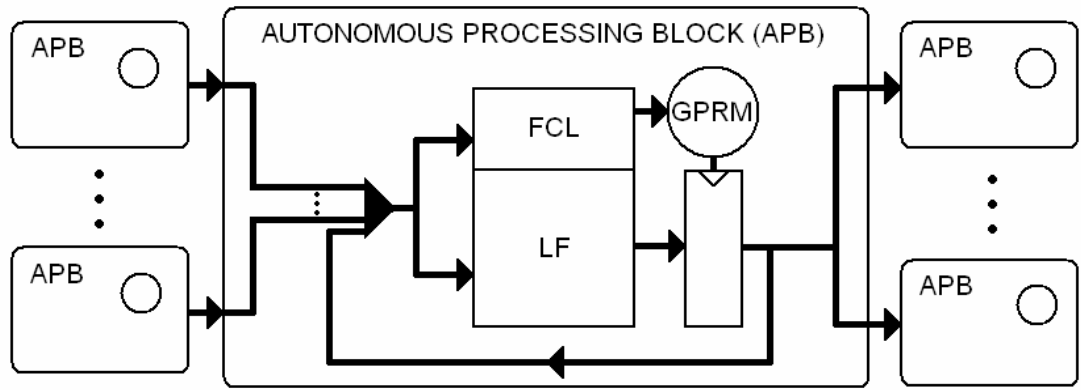

**Figure 1. Proposed system decomposition**

Each autonomous processing block has an associated GPRM which has the task of guaranteeing the correctness of data flow throughout the whole system by providing the signals for register actualization. To this end, we establish a link between the rendezvous modules in those APBs which can interchange data during system operation. Additionally, we must deploy an isolated link in the form of closed-loop over the GPRMs placed in those processing blocks in which the logic function has its registered value as an argument for sequential operation. Throughout these two kinds of links, the flow of a series of tokens provides all the information necessary for signaling data communication and clock generation respectively.

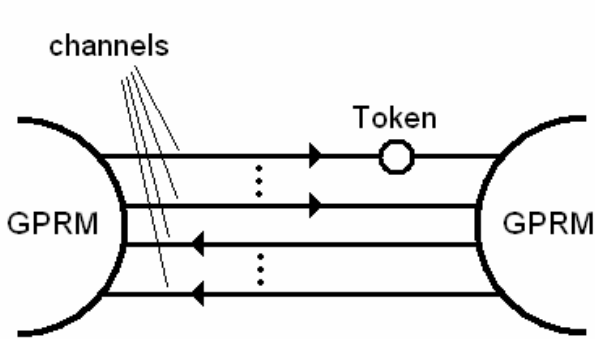
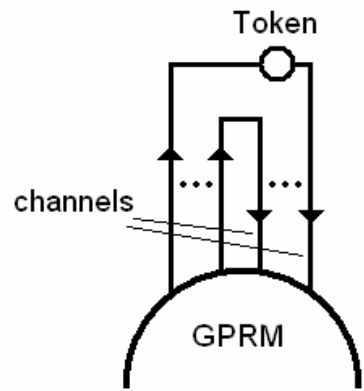

Figure 2.a. Communication signaling link    Figure 2.b. Closed-loop link

**Figure 2. Functional operation of the two links used in the protocol (a, b)**

Figure 2 shows the functional operation of the links. If we focus our attention on those links associated to data communication between two processing blocks, we can see that the respective associated GPRMs share a unique token. These links have a plurality of channels in the two directions of the communication, having each channel a different propagation delay. Now, if we look at the closed loop link, there is a unique token again as well as a set of channels with different delay characteristics. We must note that the number of channels in each direction of a communication link or in a closed-loop link is not fixed in order to introduce the most general implementation case.

The existence of multiple channels with different delays is aimed to allow deterministic selection of time spent by data transfers in communication links and clock cycle length in closed-loop links. For an efficient design, we will always use single-channel implementation and will only deploy a concrete multichannel link when we were interested in using the delay selection ability for specific purposes. In this way, by using deterministic delay selection in a smart fashion, we can control how fast our system works or even implement EMI mitigation techniques without introducing an appreciable logic overhead.

## 2.2. SIGNALING

Depending on the placement of a token over the link, the rendezvous module will interpret different messages. If the controlled link is associated to an output towards another processing block that must use the value of the register, the presence of the token indicates that destination APB has already used the stored data and has solicited its actualization. If the token is not present, the rendezvous module interprets that stored value is still being used by the destination processing block so the register must not be updated.

If the GPRM is controlling a link associated to an input of its processing block, the presence of the token indicates that the data originated by the external APB has been received and processed correctly and is ready for register actualization. On the other hand, the absence of the token indicates that the data are not ready yet.

Last, the presence of the token in the point controlled by the GPRM on a closed loop link indicates that the information stored by the register has completed the feedback datapath inside the own APB and is ready for register actualization. If the token is not present, the rendezvous module interprets that data are still being processed by the logic function.

When the GPRM detects the presence of a token in all the links that it is controlling, it interprets that every necessary data for evaluating the next value of the register has been received and processed and that all the external processing blocks that use the stored value have solicited an update. Under these conditions, the rendezvous module generates a clock pulse that updates the data stored in the register.

Prior to this instant, the GPRM must have decided which ones of the tokens are going to be sent throughout their respective links and, if doing that, which channels must be used. In order to do that, the processor block contains dedicated flow control logic (FCL) which indicates to the rendezvous module the next movement of the tokens. As we can see in Figure 1, we can consider this control logic a differentiated function parallel to the main logic function, having the same arguments, and for practical purposes we can merge both them into a global one.

Now, we are ready to take into consideration the dynamic behavior of the tokens. If we consider the case of an output associated link, when the rendezvous module decides to send the token to the destination processing block it means that it is sending the data contained in the register too. For a correct reception of the data in the destination APB, the propagation delay of the selected channel must be greater than the worst-case time that data spends in travel from the output of the origin register to the input of the destination one. This kind of signaling is known as bundled-data asynchronous protocol [5]. If the GPRM decides to keep the token, this means that the data just updated in the register is not the value which destination processor block needs.

When a GPRM decides to send back an input associated token, the meaning of that action is a data update request necessary for obtaining the next value for the register. In this case, there isn't any consideration about the delay of the chosen channel. If the rendezvous module decides to retain the token till a new register updating happens, it implies that the data present in the associated input is going to be used again by the logic function.

Finally, when the information contained by the register changes and crosses the logic function loop inside the own APB, the rendezvous module must send the token throughout the closed-loop link. In order to guarantee that the next register update will be correct, the delay of the chosen channel must be greater than the delay of the worst-case datapath in the logic loop. Note that if we design the system carefully, in most of the cases we could eliminate the closed loop link of the GPRM, but this is strictly necessary in those processing blocks that need to work in sequential mode while retaining the tokens associated to the communication links. In fact, the closed-loop link is the base of the rendezvous module ability for generating controlled clock bursts.

As we have said before, when the GPRM detects the presence of a token in each of the links that it is controlling, it generates a clock pulse. Prior to generating the next clock pulse, the rendezvous module must lose at least one of the tokens; then, when the GPRM recovers all the tokens again, a new synchronization signal will be produced. The problem arises when after generating a clock pulse the processing block doesn't send any data to destination APBs or update request to source APBs, so that all the tokens associated to communication links remain in the GPRM. In this situation, the closed-loop link assures the correct generation of clock pulses for sequential operation of the processing block till the rendezvous module uses a communication link again. For doing that, the GPRM sends the closed-loop link associated token; after crossing throughout the link, the rendezvous module recovers it again. This causes a momentary lost of the token that allows continuous clock burst generation.

## 2.3. LOGICAL IMPLEMENTATION

Next, we will introduce the logic implementation of the links and the GPRM optimized for LUT-based FPGAs. In order to match with the special architecture of theses devices, we have adopted a two-phase asynchronous handshaking protocol [3, 5] based on the propagation of parity changes.

In Figure 3.a, we show the design of a communication link between two different rendezvous modules. This is basically implemented by a set of T-type flip-flops in each side of the link. In the general case, we have as many T flip-flops as channels deployed in each direction of the link. Initially, all the flip-flops contain the "0" logic value, so the parity of the channels together is even in both sides of the link. In order to guarantee that there is a single token in the link, we place a XOR function in one side of the link and a XNOR function in the other. We will assume that when the output of one of these functions is set to "1" logic value, it means that the token is present in that side of the link. In this way, the token will be initially placed in the rendezvous module associated to the XNOR link side. Now, in Figure 3.b we show the logic implementation of the closed-loop link, which is very similar to the communication one. The main difference is that in the closed-loop link only the XNOR function is present, so the token will be always placed initially in the associated GPRM.

These kinds of links reduce the task of designing a certain rendezvous module to adding an AND function to the outputs of every XOR and XNOR functions associated to that GPRM. In fact, the rendezvous module itself is implemented over the FPGA by the resulting function mapped into look-up tables plus the sets of T flip-flops which are being controlled. Now, when all the tokens are present, the AND function output is set to "1" and it produces a rising clock edge - in order to guarantee the correct operation of the system, the AND function of those GPRMs that initially have all the tokens must be forced to "0" by an external control signal - . This clock signal propagates throughout a local clock domain and causes the updating of the registered logic function and the T flip-flops. If the fan-out of the AND function is high enough for affecting signal integrity, we must deploy a clock buffer. Is important to note that system performance will depend on the way we use these FPGA embedded clocking resources [1], but this discussion is out the scope of this paper. Once the T flip-flops have been updated, GPRM internal feedback paths force the AND output again to "0".

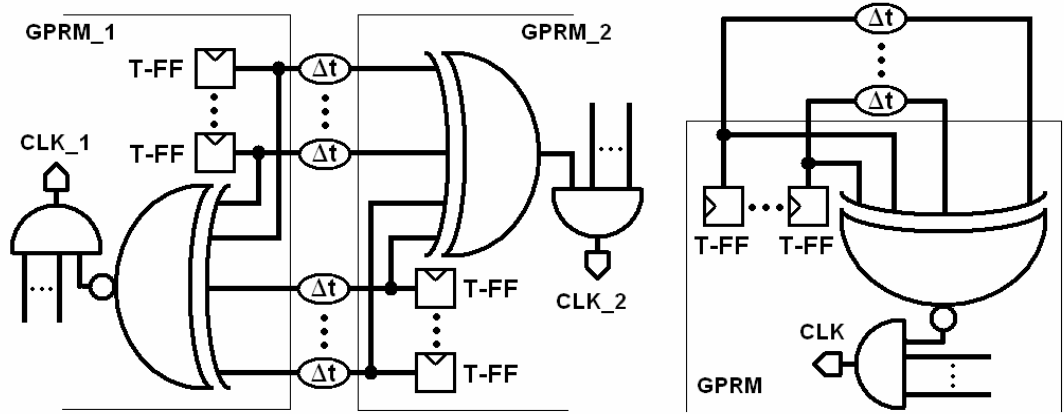

Figure 3. Logical implementation of the links over FPGAs (a, b)

Previously, not only the logic function has evaluated the new value for the register, but also the flow control logic has enabled the transition of those T flip-flops associated with the selected channels. In this way, the clock signal will cause a parity change that will propagate to the other side of the selected links with a temporal characteristics fixed by the delay unit of the involved channels signaling the desired data transfers. Note that we usually use a chain of latches for implementing delay units, but these can be deployed in many different ways. If we focus our attention in communication links, induced parity changes signalize the movement of the tokens from one side of the link to the other; in closed loop links, a transition indicates the loss of the token till the parity change crosses throughout the selected delay unit. Thanks to the special characteristics of parity signaling, these token transition processes will be valid independently of the values adopted by the T flip-flops along the time.

| GPRM FUNCTIONALITY | NUMBER OF LINKS | | | LOGIC RESOURCES | |
|---|---|---|---|---|---|
| | INPUT | OUTPUT | LOOP | 4 INPUT LUT | SLICE T-FF |
| Pipeline Stage Controller | 1 | 1 | 0 | 1 | 1 |
| Pipeline Stage Controller * | 1 | 1x2 channels | 0 | 2 | 3 |
| Pipeline Join ( 2 to 1) | 2 | 1 | 0 | 1 | 1 |
| Pipeline Fork ( 1 to 2) | 1 | 2 | 0 | 1 | 1 |
| Sequential Machine Controller | 1 | 1 | 1 | 2 | 3 |
| Sequential Machine Controller | 1 | 1 | 1x2 channels | 3 | 4 |

\* An aditional LUT input will be required in destination GPRM

**Table1. FPGA resources consumed by the GPRM**

Before ending this section, we must note that the amount of FPGA resources used by a GPRM can be strongly optimized depending on the functionality of the rendezvous module itself. Not only the logic function inside the rendezvous module is mapped optimally into LUTs, but different T flip-flops can be redundant an optimized too. We are currently exploring system partitioning techniques in which standard EDA tools are in charge of doing this optimization automatically. As way of example, in Table 1 we can see the most frequently used configurations of the GPRM and the logic resources consumed by these in Spartan3 and Virtex4 Xilinx devices. Additionally, Figure 4 shows the placement of two straight-forward pipeline stages and their associated GPRM.

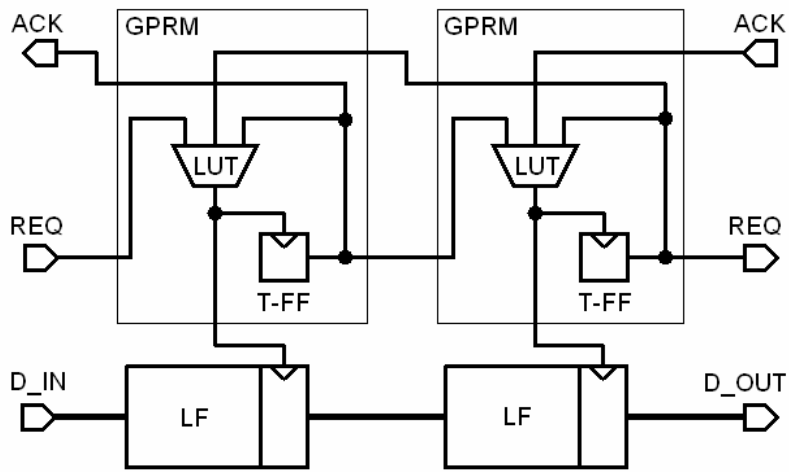

**Figure 4. FPGA implementation of two straight-forward asynchronous pipeline stages**

## 3. EXPERIMENTAL RESULTS

In this section, we will introduce some of the most interesting experimental results from the laboratory tests we have conducted. We have been working over Spartan3 and Virtex4 Xilinx devices with the ISE software and, although we have implemented a high variety of designs in order to validate our methodology [4, 6], the most of the presented empirical conclusions are derived from the behavior of FIFO and ring pipelines.

### 3.1. TIME PERFORMANCE CONCERNS

As we have seen along the previous discussion, as any GALS approach the main base of our methodology is to use conventional FPGA synchronous logic blocks for processing tasks. Then, when we have finished deploying the system, we add our specially designed 2-phase parity based asynchronous circuitry that has the full control over all the system in terms of data transfer between blocks and clock generation. In our laboratory test, we have observed that time performance of our technology is high enough to control any synchronous block implemented over the FPGA, reaching in the limit the values that we can see in Table 2 at 25ºC.

| Maximum experimental time performance at 25 °C | | |
|---|---|---|
| Device under test | Communication Link | Closed-loop Link |
| Xilinx Spartan3, -4 speed grade | > 175 MDI/s | > 300 MHz |
| Xilinx Virtex4, -4 speed grade | > 550 MDI/s | > 700 MHz |

**Table 2. Maximum speed of the links over Xilinx devices**

### 3.2. ELECTROMAGNETIC INTERFERENCE MITIGATION

The ability of our GPRMs on selecting the time evolution of data transfers throughout the system allows us to act over the power spectrum of the electromagnetic interference (EMI) that the FPGA emits in both conducted and radiated mode. In the *SuperpipeS3* test series, we analyzed the conducted EMI of a Spartan-3 Xilinx device which was programmed with a depth switched synchronous pipeline vs. the same design using our GALS approach. In this last pipeline, we implemented a spread-spectrum technique allowing to only one of the communication link to have the ability to choose between two different delays by producing a pseudo-random pattern in the associated flow control logic [7]. In Figure 5.a we can see the experimental results for conducted EMI in the synchronous pipeline while Figure 5.b shown the GALS approach ones – we can appreciate than an external FM radiation causes an induced interference at about 100 MHz-.

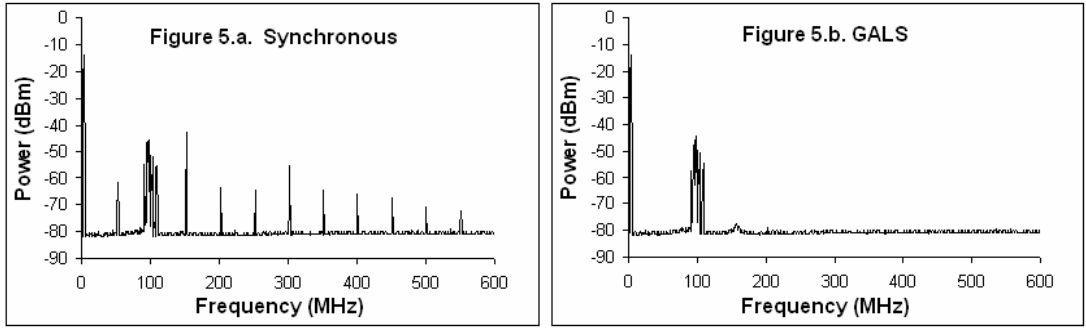

**Figure 5. Conducted EMI measured in SuperpipeS3 test series (a, b)**

## 3.3. ROBUST BEHAVIOR IN EVOLVING ENVIRONMENTS

Temporal characteristics of CMOS technology depend in some part on device operating conditions as temperature or supply voltage [8, 9]. If we focus our attention in temperature, we implement the asynchronous circuitry with the same parts than the processing blocks, so when an overheating occurs in the FPGA and causes the rise of logic delays, we expect that deployed asynchronous network slow-down adapting the time evolution to the new hardened environment. For evaluating this, we designed the *HotterV4* test series. In Figure 6, we show the variation of throughput in an extremely deep pipeline controlled by a GPRM module and implemented over a Xilinx Virtex4 when we force a cooling-fan failure. As the device temperature rises, the system automatically slows down reducing the amount of heat that the package needs to dissipate and finally the temperature and the throughput reaches a stationary state. We are currently conducting similar tests that involve automatic adaptation to supply characteristics changes.

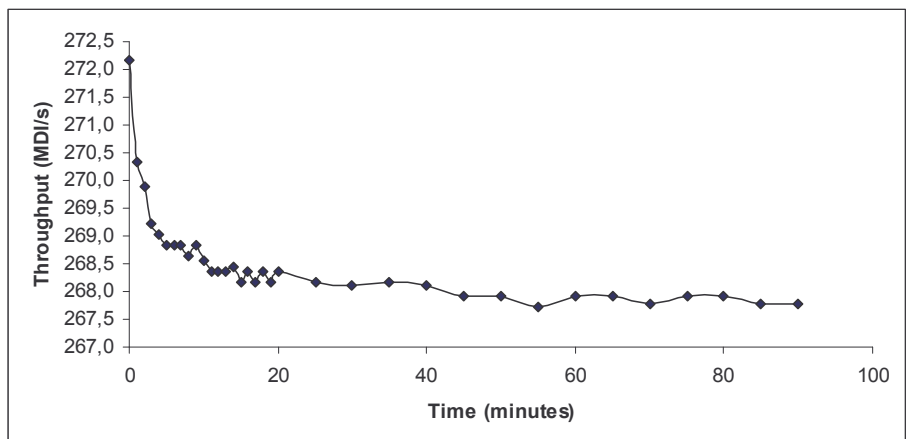

**Figure 6. Throughput auto-compensation in a cooling-fan failure test**

In those cases in which automatic adaptation to the environment is not enough, we can give the system the ability of sensing its own temperature or supply voltage. By using that information, our GALS approach can select the time performance that matches best with its operating environment avoiding the risk of failures associated to external derives.

## 4. CONCLUSION

In this paper we have firstly explained what the origin of the recent interest associated to implementing GALS systems over programmable devices is, and how conventional approaches are not capable to deal in an efficient way with the fixed architecture of readily available LUT-based FPGAs. In order to overcome this issue, we have introduced a new GALS paradigm based on the micropipeline concept and its associated control circuitry. In this way, we have proposed a system partition into autonomous processing blocks which rely on a 2-phase / bundled data parity-based handshaking protocol for communication and clock generation tasks. This new signaling protocol has smart real-time delay selection added functionality over previous ones. We have also explained an efficient logic implementation of this GALS approach over commercial FPGA architectures.

In order to validate our technology, we have shown the most interesting empirical conclusions obtained in the experimental tests that we have conducted over Spartan3 and Virtex4 Xilinx devices. We have found that our circuits widely surpass the most demanding speed requirements. Additionally, and by using the real-time delay selection ability of our circuits, we have developed efficient spread-spectrum EMI mitigation techniques and we have pointed to a new methodology that allows smart adaptation to device environment changes.